\documentclass[aps,prl,twocolumn,amsfonts,amssymb,amsmath,floatfix, superscriptaddress]{revtex4}

\usepackage{graphicx}
\usepackage{color}
\usepackage{subfigure}
\usepackage{bm}
\usepackage{epstopdf}
\usepackage{comment}
\usepackage{float}

\begin{document}
\title{Stoner ferromagnetism in a thermal pseudospin-1/2 Bose gas}


\author{Juraj Radi\'{c}}
\affiliation{Joint Quantum Institute and Condensed Matter Theory Center, Department of Physics, University of Maryland, College Park, Maryland 20742-4111, USA}

\author{Stefan S. Natu}
\affiliation{Joint Quantum Institute and Condensed Matter Theory Center, Department of Physics, University of Maryland, College Park, Maryland 20742-4111, USA}

\author{Victor Galitski}
\affiliation{Joint Quantum Institute and Condensed Matter Theory Center, Department of Physics, University of Maryland, College Park, Maryland 20742-4111, USA}
\affiliation{School of Physics, Monash Univeristy, Melbourne, Victoria 3800, Australia}

\date{\today}

\begin{abstract} 
We compute the finite-temperature phase diagram of a  pseudospin-$1/2$ Bose gas with contact interactions, using two complementary methods: the random phase approximation (RPA) and self-consistent Hartree-Fock theory.  We show that the inter-spin interactions, which break the (pseudo) spin-rotational symmetry of the Hamiltonian, generally lead to the appearance of a magnetically ordered phase at temperatures above the superfluid transition. In three dimensions, we predict a normal easy-axis/easy-plane ferromagnet  for sufficiently strong repulsive/attractive inter-species interactions respectively. The normal easy-axis ferromagnet  is the bosonic analog of Stoner ferromagnetism known in electronic systems. For the case of inter-spin attraction, we also discuss the possibility of a \textit{bosonic} analogue of the Cooper paired phase. This state is shown to significantly lose  in energy to the transverse ferromagnet in three dimensions, but is more energetically competitive in lower dimensions. Extending our calculations to  a spin-orbit-coupled Bose gas with equal Rashba and Dresselhaus-type couplings (as recently realized in experiment), we investigate the possibility of stripe ordering in the normal phase. Within our approximations however, we do not find an instability towards stripe formation, suggesting that the stripe order melts below the condensation temperature, which is consistent with the experimental observations of Ji \textit{et al.} [Ji \textit{et al.}, Nature Physics \textbf{10}, 314 (2014)].
\end{abstract}

\maketitle

The interplay between superfluidity/superconductivity and competing orders such as magnetism or density-wave ordering is one of the main challenges in the physics of strongly correlated systems, ranging from high-Tc superconductors to neutron stars. A paradigmatic system where this physics can be explored is a two-component Bose gas \cite{Cornell1998_1, Wieman1997,Chin2013}. While the zero temperature physics of binary Bose condensates (BEC) is well understood \cite{Ho1996,Cornell1998_2}, attention is turning to understanding the properties of strongly interacting binary systems which can be realized either by using Feshbach resonances \cite{Tiesinga2010}, optical lattices \cite{Greiner2002,Ketterle2009} or band engineering \cite{Chin2013}. Such systems exhibit a variety of novel phenomena such as a paramagnetic-ferromagnetic transition~\cite{Chin2013}, stripe orders \cite{Ho2011,Stringari2012}, and Mott states with residual phase coherence \cite{Svistunov2003,Altman2003}. Here we discuss the normal state properties of an interacting, pseudospin-$1/2$ Bose gas, finding a rich phase diagram, where magnetic order occurs even without superfluidity. 

\begin{figure}[h]
\centerline{
\mbox{\includegraphics[width=0.65\columnwidth]{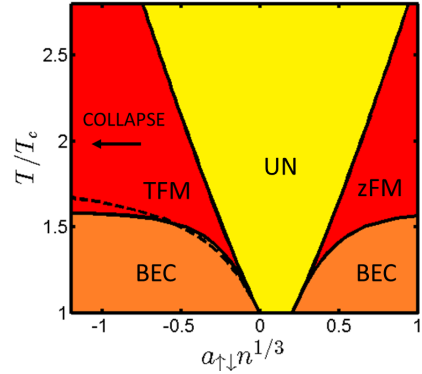}}
}
\caption{(color online)
\textbf{3D Finite-temperature phase diagram of a pseudospin-$1/2$ Bose gas--} Intra-species interactions are  repulsive ($g_{\uparrow \uparrow}=g_{\downarrow \downarrow}=g = 4\pi\hbar^{2}a/m>0$; we set $a n^{1/3}=0.1$), while the inter-species interaction $g_{\uparrow \downarrow}$ varies. 
$T_c$ is the ideal gas Bose condensation temperature (only $T>T_c$ is shown), $a_{\uparrow \downarrow}$ is the inter-species scattering length, and $n$ is the total density. For $|g_{\uparrow \downarrow}|$ greater than a critical value, system develops  ferromagnetic order in z-direction (ZFM)/$x-y$ plane (TFM) above the superfluid transition. Collapse  occurs for sufficiently large negative $g_{\uparrow\downarrow}$ (see Fig.~\ref{stability_plot}). Dashed line shows the transition between the normal unpolarized state (UN) and the paired state. TFM is always favored over pairing in $3$D. 
} 
\label{pdschematic}
\end{figure} 

Our main result is summarized in Fig.~\ref{pdschematic}, which shows the  phase diagram of a uniform pseudospin-$1/2$ Bose gas with contact interactions in three dimensions (3D) as a function of temperature ($T$) and the inter-spin interaction parameter ($g_{\uparrow\downarrow}$). This phase diagram was calculated within a self-consistent Hartree-Fock (HF) approximation, described below. Due to the synthetic nature of the spin, contact interactions generally do not preserve spin-rotational symmetry, and break it down to $U(1)\times Z_2$ in the underlying Hamiltonian. This leads to the appearance of intermediate \textit{normal} magnetic phases at finite temperature, in addition to the unpolarized normal phase (UN).  For repulsive inter-component interactions, we find an easy-axis ferromagnet in the $z$-direction (zFM), which breaks $Z_{2}$ symmetry; for attractive interactions, we predict an easy-plane transverse ferromagnet, which breaks $U(1)$ symmetry in the $x-y$ plane. 

The transition from a fully disordered phase to the zFM for strong repulsive $g_{\uparrow \downarrow}$ is reminiscent of the Stoner
transition in an itinerant electronic system (such as ultra cold Fermi gases or a screened Coulomb gas). There, the large repulsive interaction energy cost can be offset by the formation of ferromagnetic domains. Recently, itinerant ferromagnetism was investigated in strongly interacting ultra-cold Fermi gases \cite{Duine2005,Pekker2011,Ketterle2012}, which concluded that the Stoner transition is preceded by the rapid formation of bound pairs, that lead to atom loss, preventing the observation of ferromagnetism \cite{Ketterle2012}. Here we show that for the analogous bosonic system, the critical interaction strength for the onset of ferromagnetism is lower ($k_{\rm T} a_{\uparrow \downarrow} \sim 0.6$,
$k_{\rm T}=\sqrt{2m/\beta}/\hbar$, whereas in a Fermi system $k_{\rm F} a_{\uparrow \downarrow} \sim 1$,
where $k_{\rm F}$ is a Fermi momentum), which opens up the intriguing possibility of observing itinerant ferromagnetism in a \textit{Bose} gas. Importantly, in the normal state, three-body losses are strongly suppressed, and lifetimes of $\tau \sim 1$s have been observed \cite{Hadzibabic2013,Cornell2014}.

We also investigate the possibility of BCS-like pairing with attractive inter-species interactions. The study of boson pairing was originally motivated by exciton condensation in semi-conductors. Nozi\`eres and Saint James \cite{Nozieres1982} argued that such a phase is the ground state of  a spin-$1$ Bose gas under appropriate conditions. Recently, a paired phase of spin-$1$ bosons was predicted above the condensation temperature, which competes with Bose condensation \cite{Natu2011}. Here we find that ferromagnetism wins over pairing in $3$D, but pairing becomes energetically competitive in quasi-$2$D, suggesting that a stable paired phase may indeed occur. 


{\it Model and Stoner Ferromagnetism.---} We study a uniform system of pseudospin-$1/2$ bosons with 
contact interactions $\hat{H}=\hat{H}_{\rm kin} + \hat{H}_{\rm int}$:
\begin{equation}
\hat{H}_{\rm kin}=\int d\bm{r} \sum_{\sigma=\uparrow, \downarrow} \hat{\psi}^{\dagger}_{\sigma}(\bm{r})
\left( -\frac{\hbar^2}{2m}\nabla^2 - \mu_{\sigma} \right) \hat{\psi}_{\sigma}(\bm{r}),
\end{equation}
\begin{equation}
\hat{H}_{\rm int}= \int d\bm{r} \sum_{\sigma,\sigma^{\prime}=\uparrow,\downarrow} 
\frac{g_{\sigma,\sigma^{\prime}}}{2} 
\hat{\psi}^{\dagger}_{\sigma}(\bm{r}) \hat{\psi}^{\dagger}_{\sigma^{\prime}}(\bm{r})
\hat{\psi}_{\sigma^{\prime}}(\bm{r}) \hat{\psi}_{\sigma}(\bm{r}),
\end{equation}
where $m$ is atomic mass, $\mu_{\sigma}$ is the chemical potential and $g_{\sigma,\sigma^{\prime}}
=4\pi \hbar^2 a_{\sigma,\sigma^{\prime}}/m$ are
interaction coefficients ($a_{\sigma,\sigma^{\prime}}$ are the corresponding s-wave scattering lengths). 
Throughout, we assume, 
$g_{\uparrow \uparrow}=g_{\downarrow \downarrow} = g >0$. In addition to the $U(1)$ symmetry associated with $\psi_{\sigma} \rightarrow \psi_{\sigma}e^{i\theta}$, the Hamiltonian has $U(1)\times Z_{2}$ symmetry in spin space. The $Z_{2}$ symmetry can be explicitly broken by making $g_{\uparrow\uparrow} \neq g_{\downarrow\downarrow}$ or $\mu_{\uparrow} \neq \mu_{\downarrow}$. We assume a spin balanced gas, and set $\mu_{\uparrow} = \mu_{\downarrow}$. 

We obtain the instabilities of the normal state by computing the spin susceptibility within a Random Phase approximation (RPA) which includes exchange \cite{KadanoffBaym,Mueller2000,Natu2011}. An instability towards ferromagnetism is signaled by a divergence in the spin susceptibility at zero frequency and wave-vector. The susceptibility tensor reads:
{\small
\begin{equation}
\chi_{\alpha \beta, \gamma \eta} (\bm{k},t)= \frac{\theta(t)}{iV} \sum_{\bm{p},\bm{q}} 
\left\langle \left[ \hat{a}^{\dagger}_{\bm{p} \alpha}(t) \hat{a}_{\bm{k}+\bm{p} \beta}(t),
\hat{a}^{\dagger}_{\bm{q} \gamma}(0) \hat{a}_{\bm{q}-\bm{k} \eta}(0) \right] \right\rangle,
\label{chi_definition}
\end{equation}
}
where $\alpha,\beta,\gamma,\eta=(\uparrow,\downarrow)$ and $V$ is the volume.
The non-interacting susceptibility is: 
$\chi^{0}_{\alpha \beta, \gamma \eta}=\chi^0$ for $\alpha=\eta$ and $\beta=\gamma$,
$\chi^{0}_{\alpha \beta, \gamma \eta}=0$ otherwise.
In $3$D, 
$\chi^{0}(\bm{k}=0,\omega=0)=-[(m/2\pi\hbar^2)^{3/2}/\sqrt{\beta}] \text{Li}_{1/2}(e^{\beta \mu})$ \cite{Natu2011}, 
where $\beta=1/k_{\rm B} T$ and $\text{Li}_s(z)$ is the polylogarithm function of order $s$.

The interacting susceptibility ($\chi^{\rm RPA}$) in terms of
$\chi^{0}$, and the interaction matrix $V_{\alpha \beta, \gamma \eta}$ is:
\begin{equation}
\chi^{\rm RPA}_{ij,mn}=\chi^{0}_{ij,mn} + \sum_{\alpha \beta \gamma \eta} 
\chi^{0}_{ij,\alpha \beta} V_{\alpha \beta, \gamma \eta} \chi^{\rm RPA}_{\gamma \eta, mn},
\label{chi_RPA}
\end{equation}
where $V_{\alpha \beta, \gamma \eta}=g_{\uparrow \downarrow}
\delta_{\alpha,\bar{\gamma}} \delta_{\beta,\bar{\eta}}
+2g (\delta_{\alpha,1}\delta_{\beta,1}\delta_{\gamma,1}\delta_{\eta,1}
+ \delta_{\alpha,2}\delta_{\beta,2}\delta_{\gamma,2}\delta_{\eta,2})$ ($\bar{\uparrow}=\downarrow$,
$\bar{\downarrow}=\uparrow$).
We are interested in the static density and magnetization susceptibilities: 
$\chi_n=\delta n/\delta U$,
$\chi_i=\delta m_i/ \delta B_i$, ($n$ is the density, $U$ is an external potential, and $m_{i}$ and $B_i$ are components
of the magnetization and magnetic field). The RPA susceptibilities read:
\begin{equation}
\begin{split}
\chi^{\text{RPA}}_n &=\frac{2 \chi^0}{1-(2g + g_{\uparrow \downarrow}) \chi^0}, \\
\chi^{\text{RPA}}_x &=\chi^{RPA}_y=\frac{2 \chi^0}{1-g_{\uparrow \downarrow} \chi^0}, \\
\chi^{\text{RPA}}_z &=\frac{2 \chi^0}{1-(2g - g_{\uparrow \downarrow}) \chi^0}.
\end{split}
\end{equation}
The density susceptibility diverges for strong-enough attractive interactions,
$2g+g_{\uparrow \downarrow}=1/\chi^0$ ($\chi^0<0$), which marks collapse \cite{Natu2011}. Absent long-range interactions, the dominant instability in the density and spin channel occurs at $\bm{k}=0$. 

The divergence in $\chi_x$, $\chi_y$ or $\chi_z$ signals a transition to a ferromagnetic phase along the transverse or longitudinal direction respectively. The transition to an Ising ferromagnet (zFM) occurs only for sufficiently repulsive $g_{\uparrow \downarrow}$ ($g_{\uparrow \downarrow}=2g-1/\chi^0$), 
whereas the transition to a $x-y$ ferromagnet occurs for arbitrarily weak attractive $g_{\uparrow \downarrow}$ ($g_{\uparrow \downarrow}<1/\chi^0$) (Fig.~\ref{pdschematic}) \cite{Ashhab2005}. This is because the zFM has to overcome the extra repulsion from the intra-component interaction term. The TFM has recently been predicted in Rashba spin-orbit coupled bosons \cite{Kopietz2013}, but spin-orbit coupling is in fact not necessary for this phase.

The RPA analysis only yields the location of the instability lines and to obtain the complete finite temperature phase diagram, we use a self-consistent Hartree-Fock mean-field theory. We define mean-fields $n_{\sigma,\sigma^{\prime}}=\frac{1}{V} \sum_{k} 
\langle \hat{a}^{\dagger}_{\bm{k},\sigma} \hat{a}_{\bm{k},\sigma^{\prime}} \rangle \neq 0$. 
The Hartree-Fock Hamiltonian then reads: $
\hat{H}_{\rm HF}= \sum_{\bm{k}, \sigma, \sigma^{\prime}} \hat{a}^{\dagger}_{\bm{k}, \sigma} 
\mathcal{H}_{\sigma, \sigma^{\prime}}(\bm{k}) \hat{a}_{\bm{k}, \sigma^{\prime}} 
- E_0$,
%
\begin{equation}
\mathcal{H}(\bm{k})=
	\begin{pmatrix}
	\epsilon_{k}+2g n_{\uparrow} + g_{\uparrow \downarrow} n_{\downarrow} & g_{\uparrow \downarrow} n^{*}_{\uparrow \downarrow} \\
	g_{\uparrow \downarrow} n_{\uparrow \downarrow} & \epsilon_{k}+2g n_{\downarrow} + g_{\uparrow \downarrow} n_{\uparrow} 
	\end{pmatrix}.
\label{HF_spectrum}
\end{equation}
where $E_0=V \left[ g \left( n_{\uparrow}^2+n_{\downarrow}^2 \right) 
+ g_{\uparrow \downarrow} \left(n_{\uparrow} n_{\downarrow} + \vert n_{\uparrow \downarrow} \rvert^2 \right) \right]$,
$\epsilon_{k}=\hbar^2 k^2/(2m)-\mu$. 
$\hat{H}_{\rm HF}$ can be easily diagonalized: $\hat{H}_{\rm HF}=\sum_{\bm{k},j} 
E_j(\bm{k}) \hat{b}^{\dagger}_j(\bm{k}) \hat{b}_j(\bm{k})-E_0$, and in thermal equilibrium 
the occupation number is given by the Bose distribution $\langle \hat{b}^{\dagger}_j(k) \hat{b}_j(k) \rangle=
\left[e^{\beta E_j(k)}-1\right]^{-1}$. 
The state of the system at temperature $T$ can be obtained by finding the self-consistent mean-field 
Hamiltonian, or by minimizing the free energy of the system by varying 
$\bar{n}_{\uparrow}$, $\bar{n}_{\downarrow}$ and $\bar{n}_{\uparrow \downarrow}$.

The HF analysis predicts a second order transition to two normal ferromagnetic phases: zFM
for repulsive and TFM for attractive $g_{\uparrow \downarrow}$, at exactly the same temperatures as predicted by the RPA theory. This is not surprising because the RPA susceptibilities can be obtained by linearizing the Hartree-Fock equations of motion \cite{KadanoffBaym}.

  When the chemical potential reaches the bottom of the lower band in Eq.~(\ref{HF_spectrum}), a BEC 
transition occurs. While the critical temperature for the transition
between the unpolarized normal and BEC does not change with the interaction strength (interactions
 merely yield a constant shift to the chemical potential), 
the transition between the normal ferromagnetic and BEC phases ($T^{\prime}_{c}$) is interaction-dependent (see Fig.~\ref{pdschematic}).
This is because ferromagnetism splits the degeneracy between $\uparrow$ and $\downarrow$ in Eq.~(\ref{HF_spectrum}). In the extreme limit $g_{\uparrow \downarrow} \rightarrow \infty$, there is only one band, and $T^{\prime}_{c}$ approaches $2^{2/3} T_c$ (Fig.~\ref{pdschematic}), the critical temperature for Bose condensation of non-interacting spinless bosons. We also note that when the HF approximation is extended to the BEC
phase, it predicts the condensation transition to be first-order, which is an artifact
of the approximation \cite{Ashhab2005}. 

Although we do not expect our theory to be quantitatively accurate near the phase boundary, we believe that it correctly captures qualitative aspects of the phase diagram. Higher order terms are expected to modify the absolute value of the BEC transition temperature, it  is found to increase
in a uniform system \cite{Kashurnikov2001,Kleinert2003}, and decrease it in 
harmonic trap \cite{Hadzibabic2011}. Analogous to the case of the usual Stoner transition in itinerant fermions, fluctuations may also raise the ferromagnetic transition temperature, and make the transition first order \cite{Duine2005}. A careful analysis of these beyond mean-field effects will be the subject of future work.



{\it Pairing.---} In analogy with spin-$1/2$ fermions, it is natural to ask whether attractive interactions between $\uparrow$ and $\downarrow$ bosonic particles could also lead to Cooper pairing. Such exotic paired states of bosons have been discussed in the context of exciton condensation in semiconductors 
\cite{Eisenstein2004,Nozieres1982}, however to date, there is no experimental evidence for such a phase. Here we look for a transition between the unpolarized normal, and paired state using
a bosonic analog of Bardeen Cooper Schrieffer (BCS) theory \cite{BruusFlensberg}. We assume a non-zero pairing field 
$\Pi_{\uparrow \downarrow}=\frac{1}{V} \sum_{\bm{k}} \langle \hat{a}_{\bm{k} \uparrow}
\hat{a}_{\bm{k} \downarrow} \rangle$~\cite{BruusFlensberg} which yields a BCS-like Hamiltonian:
$\hat{H}_{\rm p}=\sum_{\bm{k} \sigma} \epsilon_k \hat{a}^{\dagger}_{\bm{k} \sigma}
\hat{a}_{\bm{k} \sigma} + g_{\uparrow \downarrow} \left( \Pi^{*}_{\uparrow \downarrow} 
\hat{a}_{\bm{k} \uparrow} \hat{a}_{\bm{k} \downarrow} + {\rm h.c.} \right)
-V g_{\uparrow \downarrow} |\Pi_{\uparrow \downarrow}|^2$.
%
We do not explicitly include HF terms since, in the absence of ferromagnetism (or long range interactions),
these terms only produce a constant shift in energy. The pairing order parameter is given by the bosonic BCS equation:
\begin{equation}
\frac{1}{g_{\uparrow \downarrow}}=-\frac{1}{V} \sum_{\bm{k}} \left[ \frac{1}{E_{k}} 
\left(\frac{1}{e^{\beta E_k}-1} + \frac{1}{2} \right) -\frac{1}{2 \epsilon_{k}^{0}} \right],
\label{gap_equation}
\end{equation}
where $E_k=\sqrt{\epsilon_{k}^2-g_{\uparrow \downarrow}^2 |\Pi_{\uparrow \downarrow}|^2}$ and
$\epsilon_{k}^{0}=\hbar^2 k^2/2m$.
We regularize the interaction strength
$g_{\uparrow \downarrow} \rightarrow g_{\uparrow \downarrow} + (g_{\uparrow \downarrow}^{2}/V) \sum_{k<k_c} 1/(2 \epsilon_{k}^{0})$ to avoid the unphysical ultra-violet divergence.

Solving Eq.~\ref{gap_equation}, we indeed find a transition to a paired phase, but the transition temperature for pairing is lower than that for the TFM phase (Fig.~\ref{pdschematic}). For $g_{\uparrow \downarrow} \rightarrow 0^{-}$, both transition lines 
converge to $T/T_c=1-C a_{\uparrow \downarrow} n^{1/3}$, $C \approx 1.848$. To study the potential coexistence between paired and ferromagnetic phases, we perform an unrestricted Hartree-Fock Bogoliubov analysis, in which we assume both $\bar{n}_{\uparrow 
\downarrow} \neq 0$, $\Pi_{ij} \neq 0$, where $i,j=(\uparrow, \downarrow)$. However, we do not find a state which minimizes the free energy, where both ferromagnetic and pairing order parameters are simultaneously nonzero. Below we show that the possibility of pairing is strongly enhanced in lower dimensions, owing to the presence of a bound state. 


%
\begin{figure}
\centerline{
\mbox{\includegraphics[width=0.6\columnwidth]{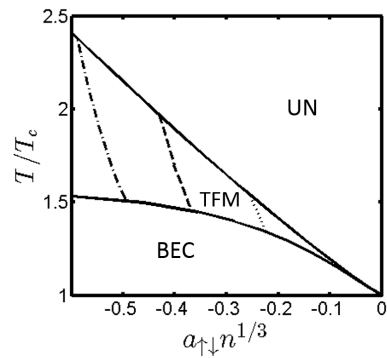}}
}
\caption{
Thermodynamic instability in $g_{\uparrow \downarrow}<0$ region in $3$D. 
UN is thermodynamically 
stable for $g>0$, while the TFM is stable in the region on the right side
of the instability lines: dotted line ($an^{1/3}=0.4$), dashed line ($an^{1/3}=0.5$)
and dashed-dotted line ($an^{1/3}=0.6$). 
} 
\label{stability_plot}
\end{figure} 

{\it  Collapse.---} As the transition to TFM occurs for attractive interactions, it is important to ask if the gas is thermodynamically stable \cite{PethicSmith}. We compute the pressure and isothermal compressibility to 
find the stable part of the phase diagram in the region with $g_{\uparrow \downarrow}<0$.
Fig.~\ref{stability_plot} shows the mechanical instability lines for different values of $a_{\uparrow \downarrow} n^{1/3}$. 
While the UN phase is stable in the entire region plotted, the TFM phase is stable only to the right of the instability lines. Increasing repulsive $g$ increases the window of stability of the TFM phase.

{\it Stoner Ferromagnetism and Pairing in $2$D.---} We now turn to the finite temperature phase diagram in quasi-$2$D, which can be realized experimentally by confinement in the $z$-direction. In quasi-$2$D, one has a new length scale, $a_z=\sqrt{\hbar/m\omega_z}$, where $\omega_z$ is the confinement 
frequency. This leads
to a momentum-dependent $2$D interaction \cite{Petrov2000,Petrov2001}:
%
$g_{\text{2D}}=\frac{2 \sqrt{2\pi} \hbar^2}{m} \frac{1}{a_z/a+(1/\sqrt{2\pi})\ln(B/\pi q^2 a_{z}^{2})}$,
%
where $q$ is the relative momentum of two particles and $B=0.915...$.

  We repeat the RPA and Hartree-Fock analysis using $g_{\text{2D}}$, and set $q=k_{\rm T}$
\cite{Petrov2001}.	
We again find zFM and TFM phases with transition
lines given by same expressions as in $3$D, however $\chi^0$ becomes:
$\chi^{0}_{2\text{D}}=-(m/2\pi\hbar^2) Li_0(e^{\beta \mu})$. In $2$D there is no 
Bose-Einstein condensation at finite temperature, however there is 
a superfluid phase below the Berezinskii-Kosterlitz-Thouless temperature ($T_{\rm BKT}$).
The approximate $T_{\rm BKT}$ for a $2$D spinless Bose gas is 
$T_{\rm BKT}/T_0=4\pi/\log(75.8 a_z/a)$ \cite{Prokofev2001}, where $T_0=\hbar^2 n/2mk_{B}$. 

We look for pairing using the same BCS mean-field approach, however with the renormalization of
interaction appropriate for a quasi-$2$D system~\cite{Salasnich2013}:
$-1/g_{2D}=\sum_{\bm{k}} (\hbar^2 k^2/m + \epsilon_{\rm B})^{-1}$, where $\epsilon_{\rm B}$
is the energy of a two-atom bound state which is related to $a_{\uparrow \downarrow}$ and $a_z$ as
$\epsilon_{\rm B}=4\hbar^2/(m a_{2D}^2 e^{2\gamma})$, where $a_{\text{2D}}=a_z (2\sqrt{\pi/B}/e^\gamma)
e^{-\sqrt{\pi/2} \ a_z/a_{\uparrow \downarrow}}$, and $\gamma=0.577...$ \cite{Petrov2001,Giorgini2011}.

\begin{figure}
\centerline{
\mbox{\includegraphics[width=0.6\columnwidth]{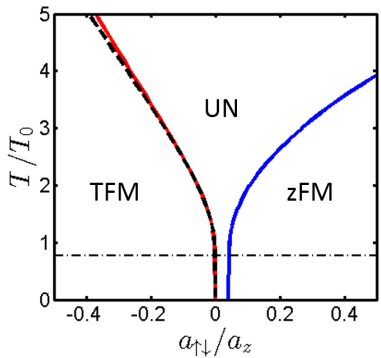}}
}
\caption{(color online)
Phase diagram in quasi-$2$D as a function of temperature and interactions
for $\eta=0.1$. We set $a/a_z=0.02$ and define $T_0=\hbar^2 n/2mk_{B}$.
Blue line shows transition between UN and zFM phase, red line shows
transition between UN and TFM phase, dashed-dotted black line corresponds to BKT transition temperature
calculated for $a_{\uparrow \downarrow}=0$. The dashed black line represents the transition between UN and 
the paired phase. 
}
\label{2D_plot}
\end{figure}

  There are four independent characteristic lengths in the system:
$a_{\uparrow \downarrow}$, $a_z$, $1/\sqrt{n}, 1/k_{\rm T}$. At fixed $a$, there are three dimensionless parameters: $a_{\uparrow \downarrow}/a_z$, $k_{\rm T}^2/n=T/T_0$ 
and $\eta=\sqrt{n} a_z$. In Fig.~\ref{2D_plot}, we show the phase diagram as a function of $a_{\uparrow \downarrow}/a_z$ and $T/T_0$ 
for $\eta=0.1$. 
 Surprisingly, unlike in $3$D, the critical temperature for transverse-ferromagnetic and paired order nearly coincide over a wide range of $a_{\uparrow\downarrow}/a_{z}$, suggesting that a stable paired phase may indeed occur in a more sophisticated treatment which includes fluctuations beyond mean-field. In particular, as the TFM and paired states are associated with $U(1)$  symmetry breaking, in $2$D, we expect vortices, which are absent in the present treatment, to play an important role. In quasi-$1$D, pairing should become even more favorable as the tendency to form bound states is much stronger in $1$D.

\textit{Experimental detection.---} In a system where the pseudospin particle number is conserved, the zFM will appear in the form of spin domains, which can either be measured \textit{in situ} \cite{Chin2013}, or using speckle imaging \cite{Ketterle2012}. The transverse components of the magnetization can be similarly obtained by using spin-echo techniques in conjunction with \textit{in situ} imaging \cite{Stamper2011}. 

The experimental realization of zFM or TFM phases in the $3$D bosonic system requires moderate interactions. As Fig.~\ref{pdschematic} shows, the FM phases should be observable for $a_{\uparrow \downarrow} n^{1/3} \sim -0.3$ ($k_T a_{\uparrow \downarrow} \sim -0.75$) for TFM, and 
$a_{\uparrow \downarrow} n^{1/3} \sim (2an^{1/3}+0.3)$ for zFM. In comparison, $^{87}$Rb has $a n^{1/3} \sim 0.02$, which means Feshbach resonances are essential for the realization of normal FM phases. However strongly interacting two-component gases can be realized using $^{85}$Rb--$^{87}$Rb mixtures \cite{Esslinger2001,Tiesinga2010}, or in $^{133}$Cs, where lattice shaking techniques can be used to create synthetic spin-$1/2$ systems \cite{Chin2013}. The situation is better in quasi-$2$D, where the region occupied by magnetic/paired phases is larger even for weak $g_{\uparrow\downarrow}$. Despite the need for moderate interactions, we stress that this physics occurs in the normal state, where three-body loss rates are significantly lower than in a degenerate gas \cite{Hadzibabic2013,Cornell2014}.

\textit{Stripe Order.---} We have generalized the RPA and Hartree-Fock theories presented above, to include the $1$D spin-orbit coupling (SOC), realized at NIST \cite{Spielman2011}. SOC splits the degenerate spin-$\uparrow$ and $\downarrow$ bands, and introduces a gap (proportional to the Raman coupling strength), and shifts the minimum of the lower band to finite wave-vectors $\pm k_{0}$, where $k_{0}$ is the wave-vector of the Raman lasers. At $T=0$, for weak Raman coupling and interactions, condensation occurs at $\pm k_{0}$, and the quantum interference of these two matter waves produces a density-wave, which spontaneously breaks translational symmetry in real space.  

It is extremely interesting to ask whether stripe order survives thermal fluctuations, and whether a normal stripe phase could occur in this system. Repeating the RPA and Hartree-Fock analyses presented above for the NIST SOC scheme, we do not find any finite wave-vector instabilities. Our negative result indicates that the stripe order melts below the transition temperature for Bose condensation, which is consistent with the experimental observations of Ji \textit{et al.} \cite{Ji2014}. Raman coupling also shifts the zFM transition to stronger interactions. This is not surprising as in the limit of large Raman coupling, the system reduces to a spinless Bose gas. 

\textit{Conclusions.---}  Observing a bosonic analog of the Stoner transition would constitute an important advance in our understanding of interacting Bose systems. Here we have established the finite temperature phase diagram of a two-component Bose gas, finding two normal, stable itinerant ferromagnetic phases, where magnetic order occurs without superfluidity. Understanding how superfluidity arises in the presence of magnetism has interesting parallels with ongoing studies of strongly correlated electronic systems. Strikingly, in $2$D, we have discussed the exciting possibility of an exotic Cooper paired state of bosons. More sophisticated calculations will be required to fully settle the question of whether pairing occurs in $2$D, and this is the subject of future work. We also concluded that there is no translational symmetry breaking in the normal phase in the continuum, consistent with recent spin-orbit coupled experiments \cite{Ji2014}. 

\textit{Acknowledgements.---} This work was supported by ARO-MURI (J.R. and S.N.), JQI-NSF-PFC (S.N.), AFOSR-MURI (S.N.), and US-ARO (V.G.). V.G. and S.N. would like to acknowledge the Aspen Center of Physics (NSF Grant No. 1066293) for its hospitality during the completion of this manuscript. We also thank Tin-Lun Ho, Erich Mueller, Arun Paramekanti, Ian Spielman, Eite Tiesinga and Shizhong Zhang for useful discussions.

\textit{Note Added:} During the completion of this manuscript, we became aware of a complementary work by Hickey and Paramekanti \cite{Paramekanti2014}. While both works discuss finite temperature aspects of spin-$1/2$ bosons, they differ in crucial ways: we \textit{analytically} solve the weak coupling, dilute limit of spin-$1/2$ Bose gases, and study transitions between the normal state and the \textit{superfluid} phase in $3$ and $2$D, whereas Ref.~\cite{Paramekanti2014} \textit{numerically} solves the strong coupling limit, and studies the transitions between the normal state and the \textit{Mott} phase in $2$D. 

\bibliography{Two_comp_references}

\end{document}